\documentclass[final,5p,times,twocolumn,sort&compress]{elsarticle}
\usepackage[colorlinks,linkcolor=red,citecolor=blue]{hyperref}
\usepackage{hyperref}
\usepackage{amssymb}
\usepackage{bm}
\usepackage{dcolumn}
\usepackage{amssymb}
\usepackage{amsmath}
\usepackage{graphicx}

\usepackage{epstopdf}
\usepackage{epsfig}
\usepackage{xcolor}
\usepackage{color}
\usepackage{subfig}
\usepackage{booktabs}
\usepackage{multirow}
\usepackage{setspace}
\usepackage{array}
\usepackage{threeparttable}
\usepackage{txfonts}

\usepackage{exscale}
\usepackage{relsize}

\makeatletter

\newcommand{\Rmnum}[1]{\expandafter\@slowromancap\romannumeral #1@}
\makeatother

\begin{document}

\journal{ Physics Letters B
}

\begin{frontmatter}

  \newcommand*{\ZZU}{School of Physics, Zhengzhou University, Zhengzhou 450001, China}
  \newcommand*{\IMP}{Key Laboratory of High Precision Nuclear Spectroscopy, Institute of Modern Physics, Chinese Academy of Sciences, Lanzhou 730000, China} 
  \newcommand*{\SNS}{School of Nuclear Science and Technology, University of Chinese Academy of Sciences, Beijing
100049, China}

\title{Probing the delicate balance of the spontaneous fission instability in sub-$\mu$s superheavy nucleus $^{252}$Rf }

  \author[a]{Zhen-Zhen Zhang}
  \author[a]{Hua-Lei Wang\corref{cor1}}
  \ead{wanghualei@zzu.edu.cn}
  \cortext[cor1]{Corresponding author.}
  \author[a]{Kui Xiao}
  \author[b,c]{Min-Liang Liu}

  \address[a]{\ZZU}
  \address[b]{\IMP}
  \address[c]{\SNS}

\begin{abstract}
Stimulated by the recent experimental discovery of the sub-$\mu$s fission nucleus $^{252}$Rf [Phys. Rev. Lett. 134 (2025) 022501], we perform an improved configuration-constrained potential-energy-surface calculation, revealing the  mechanism of intricate balance for the enhanced stability due to the high-$K$ (e.g., $K^\pi = 6^+$) isomer, possibly building on a shape isomeric state. 
The different deformation and coupling effects, such as triaxial $\gamma$, reflection-asymmetric $\beta_{3}$ and high-order $\beta_{6}$ deformations,  are discussed for both ground state and isomeric state based on the corresponding potential-energy curves along the fission valley. In particular, it is pointed out for the first time that possible multipath decay, e.g., from the high-$K$ isomeric state to those states formed between potential energy surfaces of this isomeric state and the ground state during the fission process, may reduce the nuclear lifetime and balance the fission stability. These results elucidate not only the enhanced stability of the high-$K$ isomeric state, including the inversion of stability between it and the ground state, but also the limitation of the stability increase of such an isomeric state.
\end{abstract}

\begin{keyword}
    superheavy nucleus \sep
    configuration-constrained potential-energy-surface calculation\sep 
    high-$K$ isomer \sep 
    shape isomer \sep
    enhanced stability 
\end{keyword}
\end{frontmatter}

\section{Introduction}
For a long period, the quest for new elements and nuclides at the limits of atomic number and mass has aroused great interest and the borders of the periodic table of the elements and of the chart of nuclides are still not set in stone~\cite{Smits2024}.
According to the macroscopic liquid drop model, nuclei with $Z \geq 104$ (usually defined as “superheavy nuclei”) would immediately fission due to the massive Coulomb repulsion between protons resulting in the vanishing fission barrier~\cite{Bohr1939}. At this circumstance, nuclear stability against spontaneous fission is provided only by microscopic effects of few MeV resulting from nuclear deformation, nuclear pairing, and quantum shell structure~\cite{Nilsson1969}. Indeed, the exploration of nuclear structure and stability has long been one of the main topics in nuclear physics, particularly in the superheavy and drip-line mass region. In particular, the developments of state-of-the-art radioactive-ion-beam facilities, e.g., FAIR (Germany), FRIB (USA), TRIUMF (Canada), ISOLDE (CERN), GANIL (France), SPES (Italy), NICA (Russia), RIBF (Japan), HIRFL (China), and RAON (Korea), see Ref.~\cite{Khasanov2025} and reference therein for a global review, and several parallel theories primarily including phenomenological and self-consistent mean-field models, e.g., macroscopic-microscopic models~\cite{Brack1972,Moller1994,Yang2022,Yang2022s}, Hartree-Fock methods~\cite{Goriely2009a,Goriely2009b,Ahmad2013,Blum1994} and relativistic mean-filed methods~\cite{Rufa1988,Reinhard1989,Blum1994}, and even data-driven machine-learning methods~\cite{Liu2025,Guo2025a,Carleo2019}, have renewed the enthusiasm of scientific research and greatly broadened the ability for predicting the nuclear properties of high-$Z$ superheavy elements and nuclei far from $\beta$-stability line and synthesizing them. One of the major challenges of realizing the desire to extend the nuclear chart is the smaller and smaller production rates and shorter and shorter lifetimes of nuclei as increasing $Z$ and moving away from the $\beta$-stability line~\cite{Smits2024}.

Fortunately, it was revealed that nuclear isomers, particularly the longer-lived ones, could provide an alternative and effective pathway for synthesizing, detecting and identifying shorter-lived nuclei~\cite{Xu2004,Khuyagbaatar2025}. 
Shape isomers ~\cite{Alkhomashi2009,Walker2006}, spin isomers~\cite{Xu2000,Dracoulis2006} and $K$ isomers~\cite{Jeppesen2009,Xu2004,Liu2011,liu2015} are generally regarded as three types of typical nuclear isomers according to the underlying mechanisms. 
There is no place in this letter to provide a review of nuclear isomer but the interested reader could find more details, e.g., cf. Refs.~\cite{Jain2021, Leoni2024,Jain2015} and references therein. Noted that, in the superheavy nuclei, it was pointed out that, relative to the ground state, the high-$K$ isomer can enhance the stability against spontaneous fission due to the increased height and/or widths of the fission barrier and even the stability inversion might occur between them~\cite{Xu2004}. Typical examples can be found, e.g., in $^{250}_{102}$No$_{148}$~\cite{Peterson2006}, $^{270}_{110}$Ds$_{160}$~\cite{Hofmann2001} and $^{254}_{104}$Rf$_{150}$~\cite{David2015}. In addition, it was known that, in the actinide region, the elongated shape isomer residing in the middle of the potential energy curve with a double-humped fission barrier (between the inner and outer barriers) can coexist alongside the ground state with the normally deformed shape~\cite{Bjørnholm1980}. Theoretical calculations demonstrate that the potential property with such a double-humped barrier structure was extended to the superheavy mass region and the ground state might further evolve from the normally deformed minimum to the superdeformed one~\cite{Ren2001,Ren2002,Wang2012,Guo2025}, and then the normally deformed minimum will play a role of a shape isomer.

It is important to further probe the underlying mechanism of nuclear enhanced fission-stability, originating from, e.g., the high-$K$ isomeric state possibly building on the shape isomer. Taking the recent discovery of the sub-$\mu$s fission nucleus $^{252}_{104}$Rf$_{148}$~\cite{Khuyagbaatar2025} as a carrier, we tend to reveal the effects of different deformation degrees of freedom on the fission process and understand the inversion of stability between the ground state and the high-$K$ isomeric state, especially elucidating how the expected long-lifetime of $K$ isomer is restricted. Also, we will test and briefly discuss the orbital tracking methods based on the different similarity-measure techniques in the configuration-constrained potential energy surface calculations. Prior to this work, along the fission pathway, different deformation effects, such as triaxial $\gamma$, reflection asymmetric $\beta_3$ and high-order deformation $\beta_6$, were investigated in the selected deformation spaces including generally part of them~\cite{Liu2011,Patyk1991,Wang2012,Moller2016,Chai2018,Wang2022,Liu2011b,He2019,Liu2012,Guo2025}. In the present letter, we formulate a deformation space to simultaneously cover these interested deformation parameters.

\section{Theoretical Method}
\label{Methods}
The configuration-constrained potential-energy-surface (PES) calculation is performed in an interested multidimensional deformation space ($\beta_2$, $\gamma$, $\beta_3$, $\beta_4$, $\beta_6$), allowing the simultaneous breaking of axial and reflection symmetries. We describe nuclear shape by means of the parameterization of the nuclear surface with the help of spherical harmonics and calculate nuclear potential energy within the framework of macroscopic-microscopic Strutinsky method with a realistic phenomenological mean-field Hamiltonian~\cite{Brack1972,Xu1998,Meng2022,Meng2022a}.  The microscopic single-particle levels are obtained from the deformed Woods-Saxon potential with the set of universal parameters~\citep{Cwiok1987}. We treat the pairing by using the approximate particle-number conserved Lipkin-Nagami method~\cite{Pradhan1973}, which can effectively aviod the spurious pairing phase transition encountered in the usual BCS approach. During the practical calculation, the monopole pairing is considered and the pairing strength $G$ is determined by the average-gap method~\cite{Moller1992}. The empirical windows containing dozens of single-particle orbitals, e.g., half of the nucleon number $Z$ and $N$, above and below the Fermi surface, are adopted for accommodating protons and neutrons due to the pair scattering.
The unpaired nucleon orbitals, which specify a given configuration, are tracked and blocked at each point of the deformation lattice. Several frequently-used ways (including adiabatic and diabatic ones) of identifying the configuration are briefly reviewed and tested below. For a constrained configuration, the total energy consists of a macroscopic part, which is obtained here from the standard liquid-drop model~\cite{Myers1966}, and a microscopic part resulting from the Strutinsky shell correction~\cite{Strutinsky1967}, namely, $\delta E_{\text{shell}}=E_{\text{LN}}-\tilde{E}_{\text{Strut}}$. At this moment, such a microscopic part with the inclusion of blocking effects, as discussed in the literature, e.g., see Ref.~\cite{Gaamouci2021}, will be equivalent to the summation of pairing correlation, $E_{\text{LN}}(\Delta \neq 0) - E_{\text{LN}}(\Delta = 0)$, and shell correction, $E_{\text{LN}}(\Delta = 0) - \tilde{E}_{\text{Strut}}$~\cite{Pradhan1973,Moller1995}. The configuration-constrained PES calculation can self-consistently treat the deformation, pairing and energy and, so far, has become a powerful tool in the study of multi-quasiparticle $K$ isomeric states~\cite{Shi2010, Xu1998}. This method has been applied to investigate high-$K$ fission isomers in the second well of actinide nuclei where reflection-asymmetric degree of freedom is important for the proper description of the fission barriers~\cite{Liu2011b}.   

\section{Results and Discussion}
\label{results}



\begin{figure}[htbp]
\centering
\includegraphics[width=0.48\textwidth]{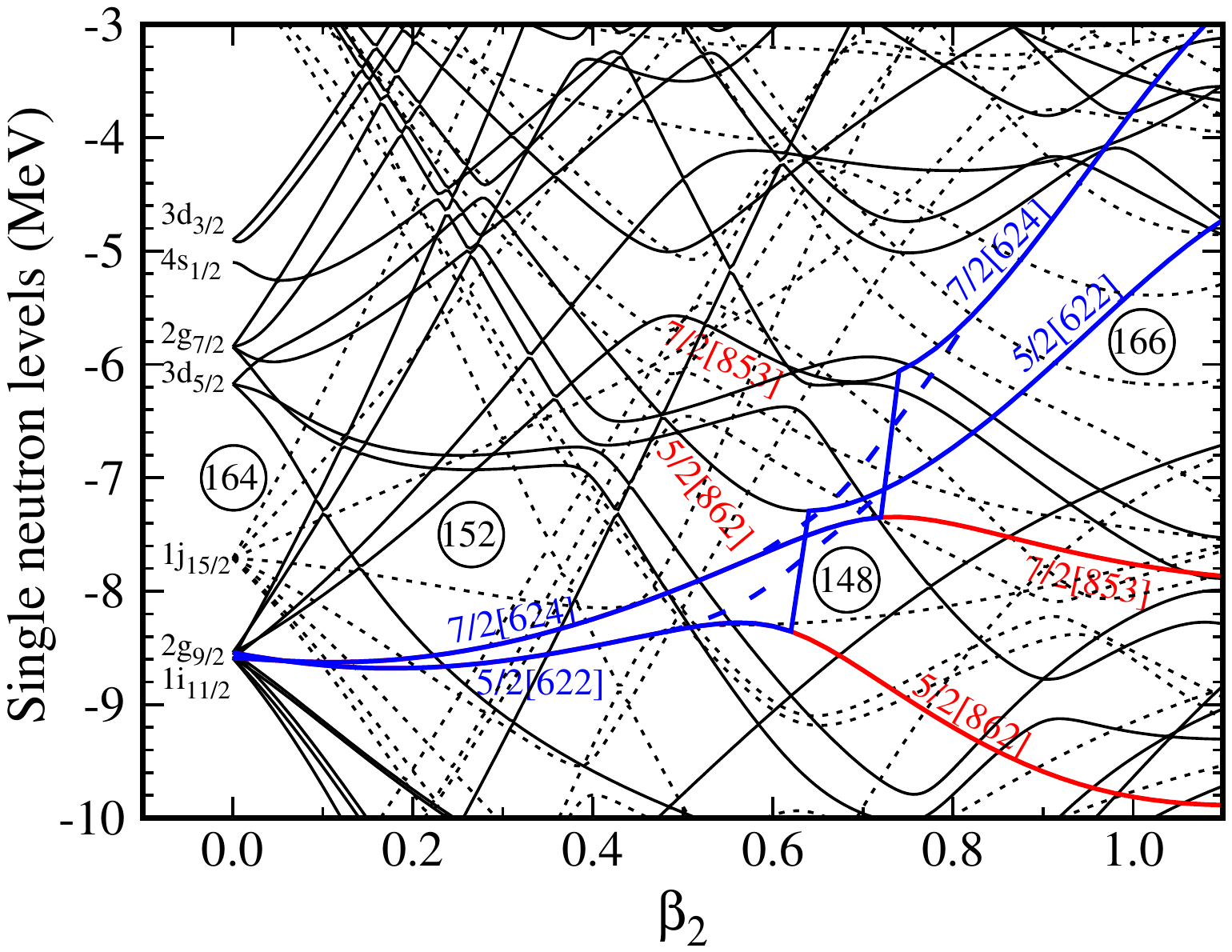}	
\caption{
Neutron single-particle energies as a function of quadrupole deformation $\beta_{2}$ for $^{252}_{104}$Rf$_{148}$, focusing on the domain near the Fermi surface. Solid and dotted curves refer to the levels with positive and negative parity, respectively. The diagonalisations of the Hamiltonian matrices for positive and negative parities are independently performed and the curves with the same parity connect energies according to non-crossing rule, except for the two interested orbitals $5/2^{+}[622]$ and $7/2^{+}[624]$. More details, see the text.
} 
		                                                           \label{fig01}

\centering
\includegraphics[width=0.45\textwidth]{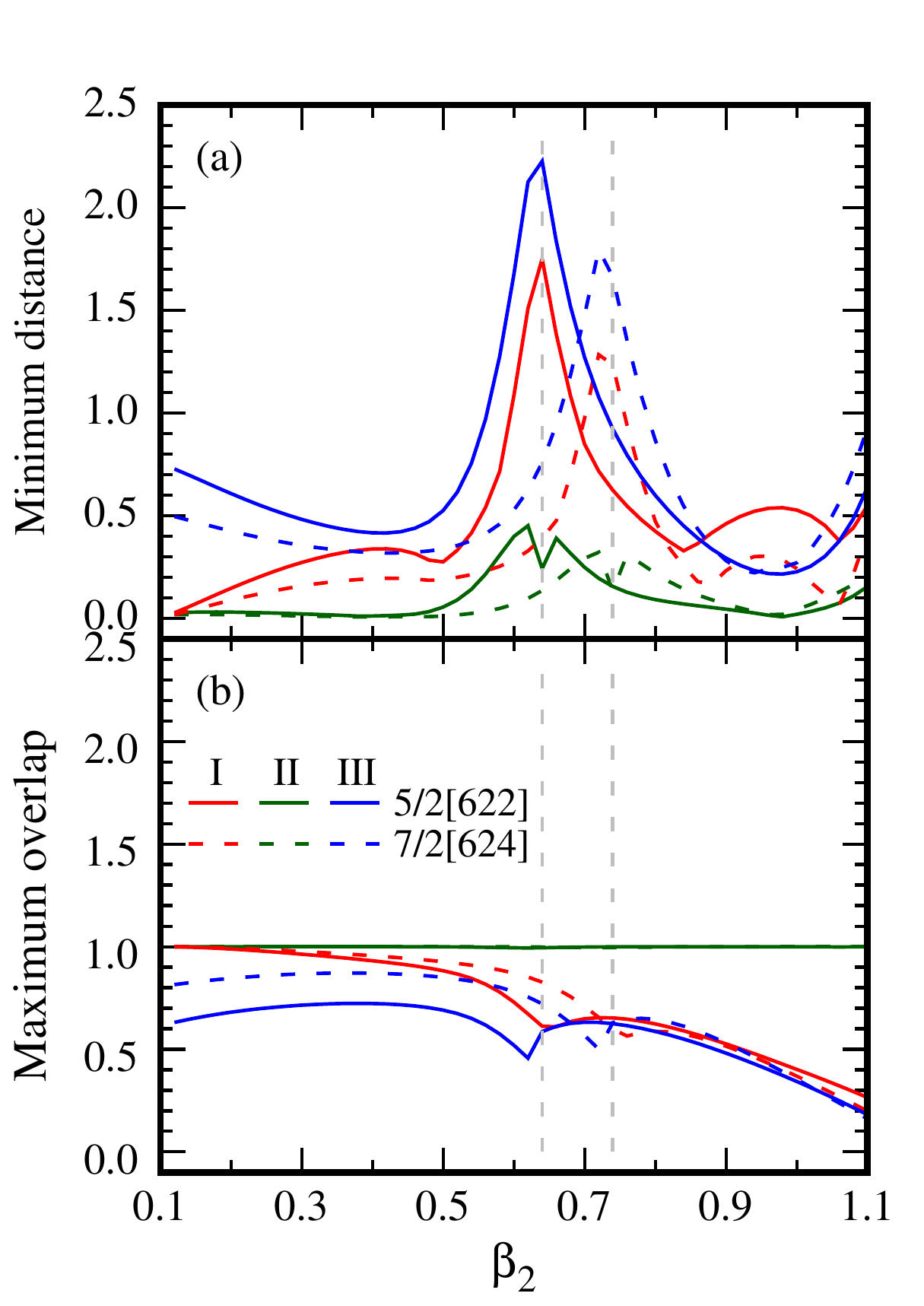}
\caption{
The minimum distance between average Nilsson quantum numbers (a) and the maximum overlap between eigenfunctions (b) as functions of $\beta_{2}$ as tracking the $\nu5/2^{+}[622]$ and $\nu7/2^{+}[624]\}$ orbitals. Solid and dash curves correspond to the $\nu5/2^{+}[622]$ and $\nu7/2^{+}[624]\}$ orbitals, respectively. Three types of color curves I (red), II (green) and III (blue) respectively correspond to three different tracking results, namely, always following the initial state, the adjacent state and the ideal state. See the text for more explanations.   
}
		                                                           \label{fig02} 
\end{figure}

Based on the PES calculations, we obtain the first energy minimum at the equilibrium deformation grid (0.24, $0^\circ$, 0.00, 0.02, $-0.02$) in the multidimensional deformation space ($\beta_2$, $\gamma$, $\beta_3$, $\beta_4$, $\beta_6$), agreeing with, e.g., the calculations by M\"{o}ller et al~\cite{Moller2016}.  
To illustrate the configuration constraints, Figure~\ref{fig01} shows neutron single-particle levels near the Fermi surface in function of $\beta_2$ for $^{252}$Rf, ignoring other deformation degrees of freedom for the sake of simplicity. At $\beta_2 = 0.00$, the single-particle levels are labeled by using the spherical quantum numbers $n$, $l$ and $j$. For the elongated shape with nonzero $\beta_2$, the spherical single-particle orbit $nlj$ will split into $j\pm1/2$ two-fold degenerate ones due to the existence of time-reversal symmetry (namely, Kramers degeneracy). Such several deformed orbitals interested here are marked with red and blue colors, along with the asymptotic Nilsson quantum numbers $\Omega^\pi[Nn_z\Lambda]$. The nuclear configurations, e.g., the low-lying high-$K$ states, can usually be described by always occupying (often called ``blocking'') the specific nucleon orbitals near the Fermi surface.

As one can see in Fig.~\ref{fig01}, the expected $K^\pi = 6^+$ isomeric state could be formed by blocking the single-neutron orbitals $5/2^{+}[622]$ and $7/2^{+}[624]$ near the equilibrium shape (e.g., $\beta_2 \approx 0.24$).
It should be noticed that with increasing $\beta_2$, according to the non-crossing rule, sometimes called the Landau-Zener effect~\cite{Zener1932,Landau1932,Rubbmark1981}, the so-called avoided crossing phenomenon occurs at $\beta_2 \approx 0.63$ between two single-particle orbitals, labeled by asymptotic quantum numbers $5/2^{+}[622]$ and $5/2^{+}[862]$ (similarly for $7/2^{+}[624]$ and $7/2^{+}[853]$ but at $\beta_2 \approx 0.74$). According to the favourable energy, the configuration of $K^\pi = 6^+$ excited state should be $5/2^{+}[622] \otimes 7/2^{+}[624]$ before the avoided crossing point $\beta_2 = 0.63$, $5/2^{+}[862] \otimes 7/2^{+}[624]$ between the avoided crossing points $\beta_2 = 0.63$ and 0.74, and $5/2^{+}[862] \otimes 7/2^{+}[853]$ after the avoided crossing point $\beta_2 = 0.74$. Note that the orbital with a continuous energy (e.g., cf. $5/2^{+}[622]$) possesses different wave-function structures before and after the avoided crossing point. For instance, the first three main components of the wave function are $\{27\%, 5/2^+[622]; 23\%,5/2^+[842]; 13\%,5/2^+[862]\}$ for the orbital $5/2^{+}[622]$ (labelled by the largest component) with single-particle energy $-8.31$ MeV  at $\beta_2=0.60$, while these components become $\{ 25\%, 5/2^+[862]; 22\%, 5/2^+[842]; 13\%, 5/2^+[622] \}$ for this energy-continuous orbital at $\beta_2 = 0.65$ (after the avoided crossing point $\beta_2 \approx 0.63$). At this moment, the largest component is changed to $5/2^+[862]$ and the orbital is labelled by $5/2^+[862]$ with red color as seen in Fig.~\ref{fig01}. Meanwhile, the corresponding components of the ``initial'' $5/2^+[862]$ state evolve from $\{36\%, 5/2^+ [862]; 25\%, 5/2^+ [622]; 10\%, 5/2^+ [\text{X}82] \}$ (here, 
the symbol X represents the Roman numeral 10) at $\beta_2 =0.60$ to $\{36\%, 5/2^+ [622]; 23\%, 5/2^+ [862]; 7\%, 5/2^+ [822] \}$ at $\beta_2 = 0.65$. Similarly, we take the largest component $5/2^+ [622]$ as the label of the ``initial'' orbit $5/2^+[862]$ after the avoided crossing point.

During the calculation of the configuration-constrained PES, e.g., for the $K^{\pi}=6^{+}(\nu5/2^{+}[622]$ $\otimes$ $\nu7/2^{+}[624])$ isomeric state, we need to track and block the $\nu 5/2^{+}[622]$ and $\nu7/2^{+}[624]$ single-particle states by identifying their similarities with the change of some continuous parameter(s) (e.g., the deformation $\beta_2$ in Fig.~\ref{fig01}). In quantum mechanics, a complete description of the state of a quantum object (or system) can be given mathematically by the state vector (or the wave function). Several types of methods on similarity measure have so far been developed, such as cosine similarity and distance-based methods~\cite{Carleo2019}. Cosine similarity method can measure the similarity between two vectors in an inner product space by calculating the cosine of the angle between them. Distance measures are used to measure the similarity between two or more vectors in multi-dimensional space and the most popular distance metrics include such as Euclidean distance, Manhattan distance and generalized Minkowski distance~\cite{Abdullah2021}. 

Figure~\ref{fig02} illustrates the tracking results of the neutron orbits $5/2^{+}[622]$ and $7/2^{+}[624]$ in function of $\beta_2$ by using the minimum distance and maximum overlap methods. The more pronounced bump and dent structure in the continuum distance and overlap values are the signals of the level virtual crossing.
In this distance-based similarity measure, as seen in Fig.~\ref{fig01}(a), we use the Manhattan distance as the performance metric which is calculated in a four-dimensional renormalization space formed by the average Nilsson  quantum numbers~\cite{Xu1998}. In the process of tracking the constrained configuration, we identify the configuration based on the minimal Manhattan distance between the reference configuration and the one at the point that we need to track. Here, we select the reference configuration by the following three ways. For instance, the single-particle wave function (the vector in the selected basis space) at the initially given deformation grid or at the adjacent deformation grid, even the ideal single-particle state (e.g., pure $5/2^{+}[622]$ and $7/2^{+}[624]$ states), could be taken as the reference state to test the configuration constraint. Based on the three selection ways of the reference configuration, the calculated minimum distances are illustrated for the two-quasineutron $K^\pi = 6^+(\nu5/2^{+}[622]$ $\otimes$ $\nu7/2^{+}[624]$ isomeric state in Fig.~\ref{fig02}(a). Similar to Fig.~\ref{fig02}(a), Figure~\ref{fig02}(b) shows the tracking results but based on the maximum overlap of wave functions which can be regarded as a variant of cosine similarity measure, e.g., cf. Ref.~\cite{Bengtsson1989}. It is found that, except for the case II in Fig.~\ref{fig02}(b) (which tracks the configuration based on the maximum overlap of wave functions between two adjacent deformation points), other five cases can follow the so-called diabatic orbitals with the similar wave-function structures, i.e., see the blue lines in Fig.~\ref{fig01}. The tracking way by case II in Fig.~\ref{fig02}(b) will follow the adiabatic orbitals after the avoided crossing points (see the red lines in Fig.~\ref{fig01}).

\begin{figure}[htbp]
	\centering
	\includegraphics[width=0.48\textwidth]{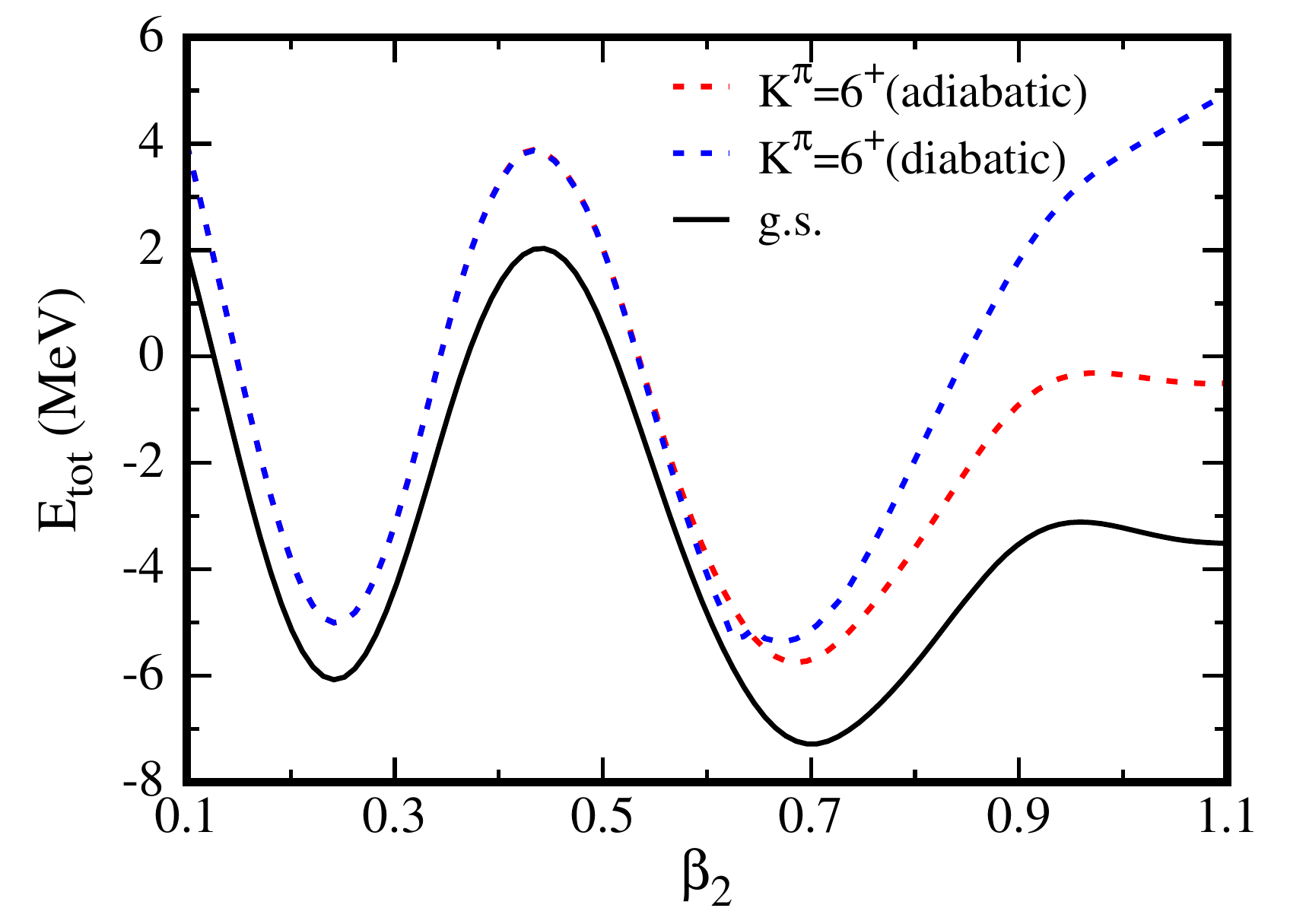}\\
	\caption{
		Calculated potential-energy curves against the
		quadrupole deformation $\beta_{2}$ for the ground state and the $K^\pi = 6^+ \{\nu5/2^{+}[622]$ $\otimes$ $\nu7/2^{+}[624]\}$ isomeric state (including the adiabatic- and diabatic-blocking results) of $^{252}$Rf. For simplicity, other deformation degrees of freedom are ignored. 
	}
		                                                           \label{fig03}
\end{figure}

To obtain the smoothly configuration-constrained PES for the diabatic calculation, we remove the virtual interaction between two avoided crossing levels, and connect the orbitals with the similar wave-function structure by e.g., a spline interpolated method (see the dash blue lines in Fig.~\ref{fig01}). Figure~\ref{fig03} illustrates the calculated potential-energy curves by blocking the single-particle orbitals $5/2^{+}[622]$ and $7/2^{+}[624]$ for the two-quasineutron $K^\pi = 6^+$ state in adiabatic and diabatic ways, together with the ground-state one for comparison. Note that during the calculation, we adopt the minimum distance method, taking the adjacent configuration as reference, to track the given configuration. One can see from this figure that, as expected, after the avoided crossing point, the adiabatic-blocking curve has the lower configuration energy but different configuration structure $5/2^{+}[862]$ $\otimes$ $7/2^{+}[853]$ though its $K^\pi$ value is still $ 6^+$. To follow the nucleon configuration with same or similar wave-function structure, the diabatic-blocking calculation is necessary, e.g., for the $K^\pi = 6^+$ state with the configuration $5/2^{+}[622] \otimes 7/2^{+}[624]$, especially along the fission path covering a large deformation domain. Nevertheless, far from the avoided crossing point, e.g., near the normally deformed minimum, both blocking ways are completely equivalent. Of course, in the practical study, a reasonable deformation space, including more deformation degrees of freedom, should be considered.    

\begin{figure}[htbp]
	\centering
	\vspace{-0.4cm}
	\includegraphics[width=0.45\textwidth]{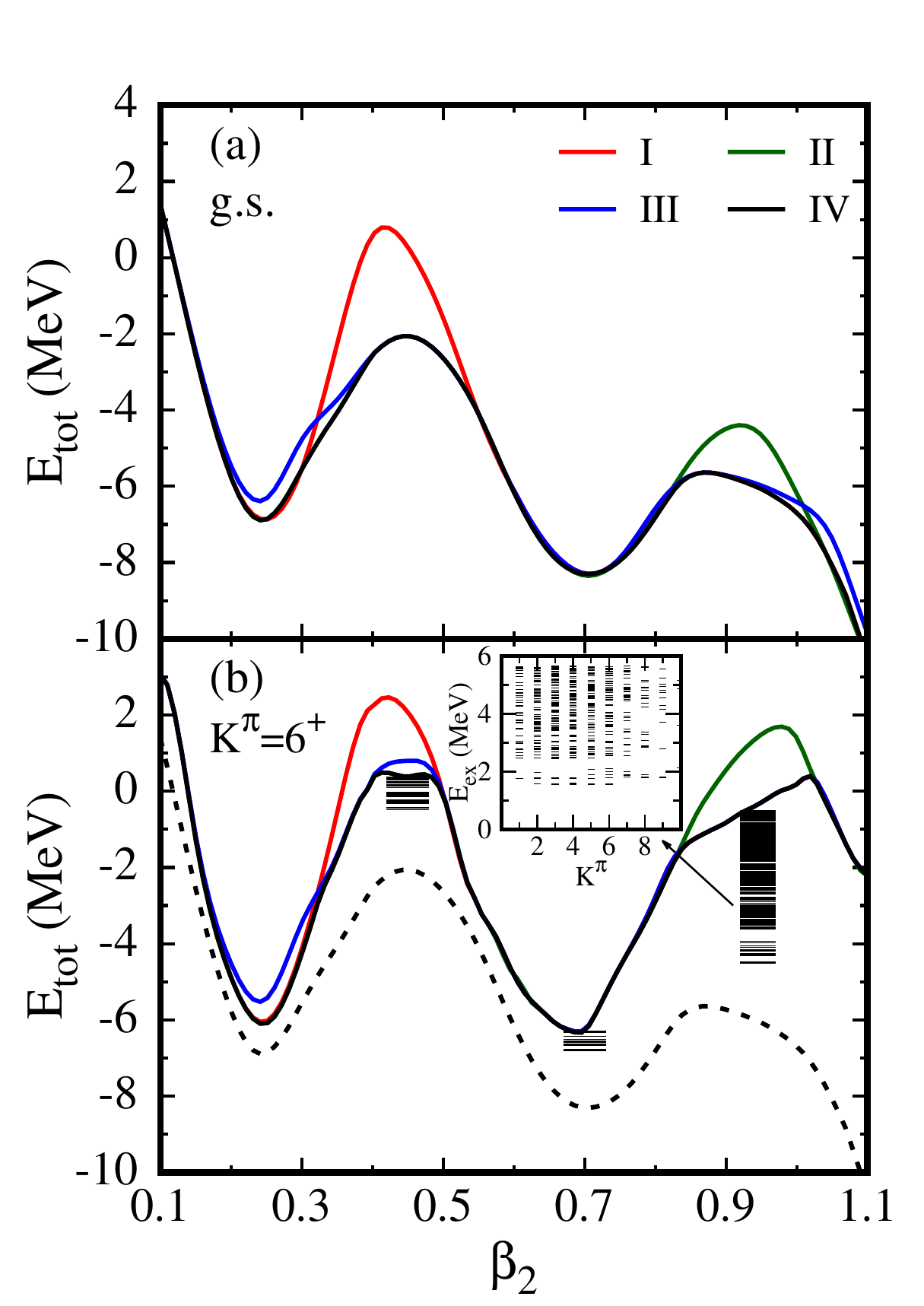}
	\caption{Four types of potential-energy curves as functions of deformation $\beta_2$ for the ground state (a) and the $K^\pi = 6^+ \{\nu5/2^{+}[622]$ $\otimes$ $\nu7/2^{+}[624]\}$ isomeric state (b) of $^{252}$Rf. Calculations are performed in the deformation space ($\beta_{2},\gamma,\beta_{3},\beta_{4},\beta_{6}$). At each $\beta_2$ point, the energy minimization is performed over \{$\beta_{3},\beta_{4},\beta_{6}$\} (I; solid red curve),  \{$\gamma,\beta_{4},\beta_{6}$\} (II; solid green curve),  \{$\gamma,\beta_{3},\beta_{4}$\} (III; solid blue curve) and  \{$\gamma,\beta_{3},\beta_{4},\beta_{6}$\} (IV; solid black curve). For comparison, the ground-state curve IV in (a) is also plotted in (b) as a dash black curve. See text for further explanations.
	}
	\label{fig04}
\end{figure}

Based on previous studies~\cite{Wang2012, Chai2018, Liu2011,Xu2024}, we select the interested deformation space ($\beta_{2},\gamma,\beta_{3},\beta_{4},\beta_{6}$) to perform the present investigation. To determine the effects of different deformation degrees of freedom on nuclear fission instability, Figure~\ref{fig04} illustrates four types of potential-energy curves with and without the configuration constraint. From this figure, one can see that the ground state, which is defined as the lowest energy configuration in a quantum system, corresponds to the superdeformed minimum ($\beta_2 \approx 0.7$ and $\approx 1.41$ MeV lower than the normally deformed minimum), agreeing with the report that some superheavy nuclei have superdeformed ground states based on the relativistic mean field calculations~\cite{Ren2001,Ren2002} and macroscopic-microscopic calculations~\cite{Wang2012,Guo2025}. However, it was claimed that, based on a macroscopic-microscopic approach, the existence of superdeformed ground states in superheavy nuclei, e.g., in $^{292}$Og, would not be supported once considering the deformation degrees of freedom of reflection asymmetry~\cite{Muntian2004}. Based on the relativistic mean-field and non-relativistic Skyrme Hartree-Fock calculations, in Ref.~\cite{Ahmad2013}, the existence of the similar superdeformed solution for superheavy nuclei was confirmed and suggested to be model-independent to a large extent. Indeed, our recent study~\cite{Guo2025} indicates that the inclusion of odd-order deformation degrees of freedom (e.g., $\beta_\lambda$, $\lambda = 3, 5$ and 7) will decrease the outer fission barrier, in agreement with the trends given in Ref.~\cite{Muntian2004}, but the minimum still may exist even if we to some extent take the model uncertainty into account, supporting the existence of superdeformed solution, even the superdeformed ground state. 

Accepting the calculated fact that the superdeformed minimum with the lower energy ($\approx -8.30$ MeV) than that (about -6.89 MeV) of the normally deformed minimum exists before the scission point, it may be more reasonable to regard the normally deformed minimum as a shape isomer. Interestingly, if so, the enhanced stability will originate from the high-$K$ ($K^{\pi} = 6^{+} \{\nu5/2^{+}[622]$ $\otimes$ $\nu7/2^{+}[624]\}$) isomeric state building on a shape isomer. Both the shape isomer and the high-$K$ one will be responsible for the enhanced stability together. In principle, the extracted barrier height and width along the fission trajectory can determine the fission half-lives in the multidimensional deformation space~\cite{Moller2009}. We investigate the impact of different deformation degrees of freedom on the fission pathway by presenting different potential-energy curves minimized over the corresponding deformation parameter(s). As illustrated in Fig.~\ref{fig04}, one can notice that, for both the ground state and the high-$K$ $K^\pi = 6^+$ isomeric state, the inclusion of the triaxial $\gamma$ and octupole deformation $\beta_3$ respectively reduces the heights and widths of the inner and outer fission barriers by affecting the corresponding saddle positions, in good agreement with our previous results (in which the high-$K$ states do not involved)~\cite{Chai2018,Wang2012}. In Ref.~\cite{Liu2013}, it was predicted that the unpaired nucleons would lead to the enhanced octupole correlation in the actinide $K$ isomeric states. At present, we do not see the octupole effect at the first minimum in the high-$K$ state of $^{252}$Rf but find that the octupole correlation has a greater impact on the right (left) part of the outer barrier at ground ($K$ isomeric) state. In the selected deformation space ($\beta_2,\gamma, \beta_{3},\beta_4,\beta_{6}$), our present calculations illustrate that the high-order $\beta_6$ deformation will affect the normally deformation minima at both ground state and $K^\pi = 6^+$ isomeric state, agreeing with the research results in Refs.~\cite{Guo2025,Liu2011,Patyk1991,Moller2016,Wang2022,Liu2011b,He2019,Liu2012}, and there is no apparent coupling between $\gamma$ and $\beta_6$ at these two cases. Recently, it was pointed out that the coupling between octupole deformation $\beta_3$ and high-order $\beta_6$ deformation may provide an interpretation of the long-standing puzzle on the rotational behavior in No isotopes~\cite{Xu2024}. Naturally, it is of interest to see, with the aid of the current opportunity, whether there exists such a coupling between $\beta_3$ and $\beta_6$ in the high-$K$ isomeric state but find nothing except for the accompanying effect between even-order and odd-order deformations near the outer barrier (also, e.g., see Ref.~\cite{Guo2025}).

From the potential-energy curves, as seen in Fig.~\ref{fig04}, relative to the cases at the ground state, the inner and outer fission barriers respectively increase by 1.65 (from 4.83 to 6.48 MeV) and 4.01 MeV (from 2.67 to 6.68 MeV) at the $K^{\pi} = 6^{+} \{\nu5/2^{+}[622]$ $\otimes$ $\nu7/2^{+}[624]\}$ isomeric state in $^{252}$Rf. Moreover, the energy difference between the normally deformed and superdeformed minima decreases from 1.41 MeV at ground state to  0.20 MeV at $K^{\pi} = 6^{+} \{\nu5/2^{+}[622]$ $\otimes$ $\nu7/2^{+}[624]\}$ isomeric state though the latter is still lower. It was pointed out that raising the barrier height by 1 MeV will increase the calculated half-life by about six orders of magnitude~\cite{Moller2009}. However, the measured half-life, $13^{+4}_{-3}\mu s$, is just two orders of magnitude longer than that of the ground-state value $60^{+90}_{-30}ns$ or so~\cite{Khuyagbaatar2025}. Obviously, such a lifetime of the high-$K$ isomeric state is not quite as long as expected, even considering a deviation between calculated and experimental fission half-lives of two to three orders of magnitude~\cite{Moller1994,Moller1976}. One of the possible explanations is numerous states might appear between the high-$K$ excited state and the ground state during the fission process. For instance, as illustrated in Fig.~\ref{fig04}(b), between the solid and dash black curves, it is found that there is no excited state at the normally deformed ($\beta_2 \approx 0.24$) minimum but many two-quasineutron excited states appear, e.g., at several selected positions, namely, $\beta_2 \approx 0.45, 0.7$ and 0.95. The inset of Fig.~\ref{fig04}(b) illustrates the detailed two-quasineutron excited states with different $K$ values (at this circumstance with $\gamma =0^\circ$, the $K$ quantum number is conserved due to the axial symmetry). Therefore, from the fact assumed above, one cannot exclude the possibility of decay of the high-$K$ isomer to these intermediate states and this scenario will lead to the stability decrease. The decay of the high-$K$ isomeric state to a rotational band is observed in $^{257}$Rf~\cite{Rissanen2013}. Of course, due to the pairing interaction, one can easily understand the existing energy gap, approaching the ground-state energy curve, in which there is no excited state. The detailed lifetime estimation for such a multipath decay process is beyond the scope of this letter but might be of interest though an accurate lifetime-calculation for spontaneous fission seems to be still difficult due to the extreme sensitivity of the theoretical half-lives to small potential-energy changes and some model uncertainties~\cite{Moller1994}. 

\section{Conclusions}
\label{conclusions}

To conclude, the spontaneous fission instability of ground state and two-quasiparticle $K^{\pi}=6^{+}$ isomeric state in $^{252}$Rf is investigated by the conﬁguration-constrained PES calculations in a formulated multidimensional deformation space ($\beta_2,\gamma, \beta_{3},\beta_4,\beta_{6}$). Both adiabatic and diabatic-blocking PES calculations are performed and compared. The mechanism of the enhanced stability, including the so-called inversion, due to both the high-$K$ and shape isomeric states is revealed and the delicate balance between stability enhancement and limitation are illustrated by proposing a new physical scheme of nuclear excitation-state decay. We also exhibit and briefly discuss the effects of different deformation degrees of freedom, such as triaxial deformation $\gamma$, reflection asymmetric deformation $\beta_3$ and high-order deformation $\beta_6$, on the width and height of the fission barrier. The findings in this project offer valuable insights into the structural properties, especially related to the possible multipath decay of high-$K$ isomeric state, at least, in the new $^{252}$Rf nuclide and its neighbours, and the refinements of corresponding nuclear models evolving the fission and synthesis probabilities in superheavy nuclei.

\section*{Declaration of competing interest}

The authors declare that they have no known competing financial
interests or personal relationships that could have appeared to
influence the work reported in this paper.

\section*{Data availability}

The data that support the findings of this study are available from the corresponding author upon reasonable request.

\section*{Acknowledgements}

This work was supported by the Natural Science Foundation of Henan Province (No. 252300421478) and the National Natural Science Foundation of China (No.11975209, No. U2032211, and No. 12075287). Some of the calculations were conducted at the National Supercomputing Center in Zhengzhou.

\section*{References}

\bibliographystyle{elsarticle-num}

\end{document}